# ARRAY BASED JAVA SOURCE CODE OBFUSCATION USING CLASSES WITH RESTRUCTURED ARRAYS

Praveen Sivadasan,
PhD Scholar, School of Computer Sciences,
Mahatma Gandhi University,
Kottayam, Kerala, India.
praveen_sivadas@yahoo.com

P Sojan Lal,
Faculty Guide, School of Computer Sciences,
Mahatma Gandhi University,
Kottayam, Kerala, India.
sojanlal@gmail.com

*Abstract— Array restructuring operations obscure arrays. Our work aims on java source code obfuscation containing arrays. Our main proposal is Classes with restructured array members and obscured member methods for setting, getting array elements and to get the length of arrays. The class method definition codes are obscured through index transformation and constant hiding. The instantiated objects of these classes are used for source code writing. A tool named JDATATRANS is developed for generating classes and to the best of our knowledge this is the first tool available for array restructuring, on java source codes.*

*Keywords—Array restructuring, Source code obfuscation, Index transformation, Constant hiding*

## 1 INTRODUCTION

Java 'byte code' retains most or all information present in the original source code [1]. This is because the translation to real machine instruction happens in the browser of the user's machine by JIT (Just-In-Time Compiler). Also, Java programs are small in size because of the vast functionalities provided by the Java standard libraries. Retaining information of the source code by byte code results in Decompilation which is is the process of generating source codes from machine codes or intermediate byte codes. Software obfuscation [2,3] is a popular and economical approach to trim down reverse engineering and re-engineering attacks through decompilation. Source code obfuscation consists of any technique that is targeted at making the source code less intelligible [10] and our focus is on obfuscating arrays of java source codes. Array restructuring operations obscure arrays by its reorganization , index, dimension and array name changes. The array restructuring operations are array splitting, array merging, array folding and array flattening [4]. Array splitting is generalized in [7] and homomorphic obfuscations [5] are to strengthen obfuscation. In array splitting, an array 'A' is split into two sub-arrays A1 and A2 where A1 holds elements of A with even indices and A2 hold elements of A with odd indices. Array merging merges two 1D arrays into one. Array folding transforms a 1D array into a 2D array and Array flattening from a 2D array into a 1D array. Analysis metrics for obfuscation have been proposed in [8 ].

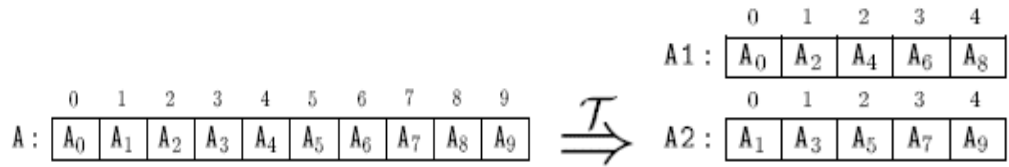

Figure 1. Array Splitting

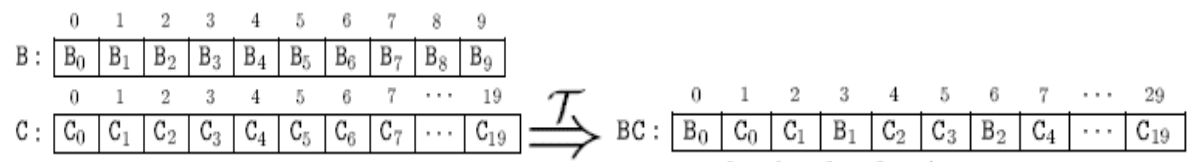

Figure 2. Array Merging

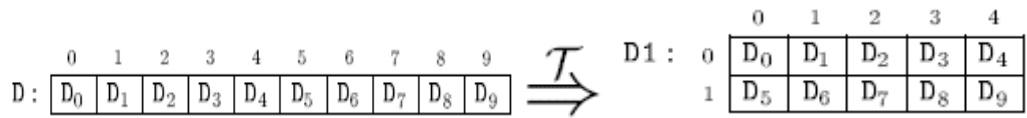

Figure 3. Array Folding

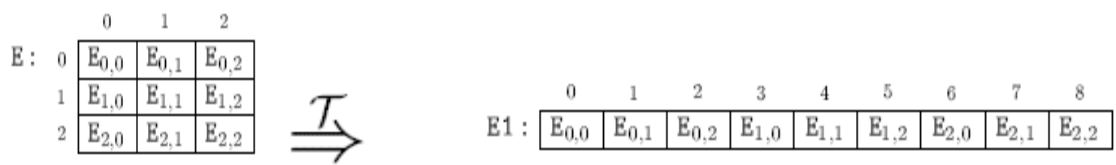

Figure 4. Array Flattening

In [6], $I = f(i) = 2 * i + 3$, is a function representing the new value of $i$ and an improvement for this index transformation is suggested in [11]. William Feng Zhu in [5] has shown a case of obfuscated program (Figure 6), for the code in Figure 5.

```
....
int A[100];
...;
S = 0; for (i = 0; i < 100; i++) S = S + A[i];
...
```

Figure 5. An unobfuscated program

```
...
int B[100];
...
S = 0; for (i = 0; i < 100; i++) S = S + B[i*3 mod 100];
...
```

Figure 6. An obfuscated program

A class say 'B' is proposed for encapsulating array object 'B [ ]' with scan and access methods. The class say 'test'(Figure 7) is written using class 'B' where the obscurity is maintained for the codes of class B, but not for class test, thus implementing partial obscurity to the entire codes.

```
class B
{        int[] B=new int[100];
         public void scan(int i, int elem) { B[i*3mod100]=elem;}
         public int access(int i){    return (B[i*3mod100]);}
}
public class test
{              public static void main(String args[])
               {             S=0; B b=new B( );
                             for (i=0;i<100;i++)  b.scan(i, 3*i + 1000);
                             for (i=0;i<100;i++)  S=S+b.access(i);
               }
}
```

Figure 7. class 'test' using class B

We also propose classes with restructured arrays. Let us consider a class 'SplitArray_Integer' with split array objects 'iObj1' and 'iObj2', in Figure 8.

```
public class  SplitArray_Integer extends obfuscate {
 int[] iObj1;int[] iObj2;
       public  SplitArray_Integer(int size )
       {     if((size%2)==0)
              { iObj1= new int[(int)(size/2)]; iObj2= new int[(int)(size/2)];}
                   else
                     {int temp=(int)(size/2)+1;iObj1= new int[temp];
                      iObj2= new int[size-temp];}}
        public void setArray(int pos,int elem){.....  }
        public int  getArray(int pos){......}
        public int  lengthArray() {......}...............}
```

Figure 8. Class with split array object members iObj1 and   iObj2

The setArray, getArray, lengthArray method definitions of class SplitArray_Integer, in Figure 9

```
public void setArray(int pos,int elem)
{      if((pos%2)==0)
              iObj1[(int)pos/2]=elem;
       else
              iObj2[(int)pos/2]=elem;
}
public  int getArray(int pos)
{  if((pos%2)==0)
       return(iObj1[(int)pos/2]);
     else
       return(iObj2[(int)pos/2]);}
public int lengthArray() {return(iObj1.length+iObj2.length);}
```

Figure 9.    Method Definitions of class 'SplitArray_Integer'

The codes of class SplitArray_Integer are obscured and this class operates only on integer arrays. This class is considered for source code writing, to instantiate 1D integer array objects (Figure 12). The other classes SplitArray_Double, SplitArray_String, SplitArray_Char are to instantiate double, string, char array objects respectively.

The method F(a,b) is used to strengthen obfuscation by  hiding the integer constant 2,[6]. The 2D array 'y-factors' stores a list of pairs of numbers whose sum gives a prime number. The values in y_factors are based on the rule that n1+n2 results in a prime number. The function calls F (41 mod 23, 2), F(6130%3071,9), F(24560%12287,11) are to hide value 2.

```
public static int F(int y,int count)
{int[][] y_factors=new int[count][2];   //A 2Dimensional array
       /*Assign values to y_factors based on the rule that n1+n2 results in a prime number*/

y_factors=[2,3],[5,6],[11,12],[23,24],[47,48],[95,96],[191,192],[383,384],[767,768],[1535,1536],
       [3071,3072],[6143, 6144], [12287, 12288];
       // Parameter value say, 2 for 'count' of  F(int,int), considers pairs of y_factors [2,3]
        ,[5,6]
        //  Parameter value say, 4 considers pairs [2,3], [5,6],[11,12],[23,24]
          for (int i=count;i>0;i--){
                 Here y1=y_factors[i-1][0]+y_factors[i-1][1];
                   y=ymody1;
          }
          return y;
}
```

Figure10. Implementation of F(int y, int count) for hiding constant 2

Also, classes with folded and flattened arrays for integer, double, string, char data types are proposed.

A tool named 'JDATATRANS' has been designed to generate the classes. The investigation of the tool is done for Splitting, Folding and Flattening operations. The key tool steps for the class generation is (i) Representing the proposed  arrays required for source code writing (ii) Decide about the array restructuring operation (iv)Generate the classes (v) Source code writing  using  the generated classes. Let us discuss a case by considering Array Splitting. The proposed arrays for array splitting are represented in a file, say 'InFile.txt' with java array object instantiation statements like

```
int[] array=new int[100000];
int[ ] a=new int[23];
double ab=new double[45];
String[ ] abc=new String[34];
char abcd[]=new char[100];
```

Figure11. Tool Input for Array Splitting

The file 'InFile.txt' is the input to the tool and generates classes SplitArray_Integer, SplitArray_Double, SplitArray_String, SplitArray_Char in the working folder, to be used for source code writing.
A source code 'test.java' written using the tool generated classes is in, Figure 11.

```
public class test {
       public static void main(String args[]){int n;
       SplitArray_Integer array=new SplitArray_Integer(100000);
       SplitArray_Integer a=new SplitArray_Integer(23);
       SplitArray_Double ab=new SplitArray_Double(45);
        SplitArray_String abc=new SplitArray_String(34);
        SplitArray_Char abcd=new SplitArray_Char(100);

       try{BufferedReader stdin =new BufferedReader(new InputStreamReader(System.in), 1);
       n=Integer.parseInt(stdin.readLine());
       for(int i=0;i<n;i++)
                  array.setArray(i,(3*i + 1000) % n);
              for(int i = 0; i <n; i++)
                  System.out.println(array.getArray(i)+"  ");

       }}
```

Figure12.  Source code written using class 'SplitArray_Integer'

An improved version  of the tool was implemented, by considering source code writing using the 12 predefined classes(Table 2). On tool invocation, '.java' files for the predefined classes, containing method definition with empty/return method statements (Figure 13) are generated. This makes to compile the source code (Figure 12) without the detailed method implementation of the class. A sample of 'SplitArray_Integer.java' files, in Figure 13.

```
public class  SplitArray_Integer  { public  SplitArray_Integer(int size ) {}
public void setArray(int pos,int elem){ }
public  int getArray(int pos){ return 0;}
public int lengthArray() {return 0;}
}
```

Figure13. Tool generated '.java' file of Class Framework in the current working folder

The next step is choosing the source code file,say test.java(Figure 12). The tool parses the source file and rewrites the'.java' files for the classes mentioned in the source code, by adding restructured arrays and detailed method implementations. The rewritten '.java' files are stored in the current working folder. The compilation of the source code file will generate the class files of the '.java' files in the present working folder, helping to execute the code.

| Sl No | InFile (No.of Arrays) | Data Types of arrays in InFile | Operations Chosen in the tool | Classes generated by the tool |
|---|---|---|---|---|
| 1 | 2 | int | | SplitArray_Integer |
| 2 | 4 | doublc | | SplitArray_Double |
| 3 | 20 | int,doublc, string,char | Array Splitting | SplitArray_Integer SplitArray_Double SplitArray_String SplitArray_Char |
| 4 | 1 | double | | FoldedArray_Double |
| 5 | 5 | int,doublc, string,char | Array Folding | FoldedArray_Integer FoldedArray_Double FoldedArray_String FoldedArray_Char |
| 6 | 10 | string | | FoldedArray_String |
| 7 | 34 | char | | FlattenedArray_Char |
| 8 | 25 | int,doublc, string,char | Array Flattening | FlattenedArray_Char |
| 9 | 73 | int | | FlattenedArray_Char |

Table 1. Tool outputs for the initial implementation

The following table shows the number of statements and function calls in the classes, for the improved version of the tool.

| Sl No | The predefined classes for source code writing | Number of statements in class with empty /return method implementation | Number of statements in each class with detailed method implementation | Number of different constant hiding function calls in each class |
|---|---|---|---|---|
| 1 2 3 4 | SplitArray_Integer SplitArray_Double SplitArray_String SplitArray_Char | 7 | 76 | 9 |
| 5 6 7 8 | FoldedArray_Integer FoldedArray_Double FoldedArray_String FoldedArray_Char | 13 | 32 | 5 |
| 9 10 11 12 | FlattenedArray_Integer FlattenedArray_Double FlattenedArray_String FlattenedArray_Char | 5 | 27 | 3 |

Table 2. The number of statements and function calls in different classes

The code 'search_orig.java'(Figure 14),search_Arraysplit.java(Figure 15),search_Arrayfold.java(Figure 16), search_Arrayflatten.java(Figure 17), are analysed for the quality of obfuscaion. The purpose of the codes is to set and get array elements.

```java
import java.io.*;
public class search_orig{
public static void main(String args[])
    {   long time = 0;int n;
        int[] array = new int[100000];
        try
        {    BufferedReader stdin =new BufferedReader(new InputStreamReader(System.in), 1);
             System.out.println("Searching.....\n");
             System.out.println("Enter the number of terms");
             n=Integer.parseInt(stdin.readLine());
             for(int i=0;i<n;i++)
                     array[i]=(3*i + 1000) % n;
             time = System.currentTimeMillis();
             for(int i = 0; i <n; i++)
                     System.out.println(array[i]+"  ");
             System.out.println("Time taken = "  + (System.currentTimeMillis() - time)/1000+ "
(Sec)");
        }catch(IOException ie){}
    }
}
```
Figure 14. search_orig.java

The main codes of search_ArraySplit.java are in Figure 15.

```
public class search_Arraysplit{
   public static void main(String args[])
   {     SplitArray_Integer array = new SplitArray_Integer(100000);
                ...................
                for(int i=0;i<n;i++)
                    array.setArray(i,(3*i + 1000) % n);
             time = System.currentTimeMillis();
             for(int i = 0; i <n; i++)
                   System.out.println(array.getArray(i)+"  ");
          ...........}}
```
Figure 15. search_Arraysplit.java

The codes of search_Arrayfold.java,Figure 16.

```
public class search_Arrayfold{
   public static void main(String args[])
   {    .................
             FoldedArray_Integer array = new FoldedArray_Integer(100000);
          ........................}}
```
Figure 16. search_Arrayfold.java

The codes of search_Arrayflatten.java,Figure 17.

```
public class search_Arrayflatten{
   public static void main(String args[])
   {     .......................
             FlattenedArray_Integer array = new FlattenedArray_Integer(500,200);
          ............
       for(int i=0;i<500;i++)
                      for(int j=0;j<200;j++)
                      array.setArray(i,j,(3*i + 1000) % 100000);
       for(int i = 0; i <500; i++)
                   for(int j = 0; j <200; j++)
                      System.out.println(array.getArray(i,j)+"  ");
       ...............................}}
```
Figure 17. search_Arrayflatten.java

Potency, resilience and cost are the factors analyzed for the quality of obfuscation [9]. The potency score,$S_{pot}$ refers to how much obscurity is added to the code that prevents human beings from understanding it. The resilience score $S_{res}$ , is a measure of how strong the program can resist an attack against a deobfuscator. The cost of obfuscation score, $S_{cst}$, refers to how much computational overhead is added to a transformed program. Due to the lack of commercial de-obfuscators in the market, this analysis is solely based on decompilation of code and $S_{res}$ could not be measured. Hence, not considering $S_{res}$, the quality of obfuscation score,$S_{quality}$ is computed using $S_{quality}$= $0.4*$ $S_{pot}$- $S_{cst}$ [8]. Since our proposal does not add Nesting, Control flow and Variable complexities to the obscured codes, weighted potency is computed by $S_{pot}$=x*$S_{LOC}$ , where $S_{LOC}$ is the ratio of how many lines of code (LOC) are added or removed in comparison to the original program length. [8], Considering the runtime score $S_{runtime}$ and storage score $S_{storage}$ ,the cost of obfuscation, is $S_{cst}$= $y_{2*}$ $S_{storage}$ + $z_{2*}$ $S_{runtime}$,where $y_2$ =0.15, $z_2$=0.45. For runtime analysis,the codes are analysed on Intel Pentium IV and Intel Core Duo machines. The runtime details are summarized in table 3,

| Source code to set and get array elements | Intel Pentium PIV,3 GHz,512 MB RAM | | | Intel Core Duo,1.66GHz,1GBRAM | | |
|---|---|---|---|---|---|---|
| | (24000 numbers) | (50000 numbers) | (100000 numbers) | (24000 numbers) | (50000 numbers) | (100000 numbers) |
| | Runtime(in seconds) | | | Runtime(in seconds) | | |
| search_orig.java | 1 | 3 | 7 | 1 | 2 | 4 |
| search_ArraySplit.java | 2 | 4 | 9 | 2 | 3 | 5 |
| search_Arrayfold.java | 1 | 3 | 7 | 1 | 2 | 4 |
| search_Arrayflatten.java | 1 | 3 | 7 | 1 | 2 | 4 |

Table 3. Runtime analysis

The LOC(Lines Of Codes of codes) for the file 'search_ArraySplit.java', is computed by considering the 22 source code statements(Figure 15), 198 statements for the 9 different function calls and the 76 statements of the class 'SplitArray_Integer'(Table 2), totaling to 296 statements.

| | LOC (Lines of Codes) | $S_{LOC}= (C'_{LOC}-C_{LOC})/C_{LOC}$ | $S_{pot} = 12.50* S_{LOC}$ | $S_{runtime}$ (100000) numbers |
|---|---|---|---|---|
| search_orig.java | 22 | 0 | 0 | 0 |
| search_ArraySplit.java | 296 | 12.45 | 155.63 | 0.2 |
| search_Arrayfold.java | 164 | 6.45 | 80.63 | 0 |
| search_Arrayflatten.java | 117 | 4.32 | 54 | 0 |

Table 4. Potency analysis

The file size $P'_{filesize}$ and $P_{filesize}$ are the sizes of the obfuscated program and original program respectively. The storage score $S_{storage}$ is computed using $S_{storage} = (P'_{filesize} - P_{filesize})/ P_{filesize}$

| | Filesize | | $S_{runtime}$ (100000) numbers | $S_{storage}$ | $S_{cst}=0.15* S_{storage} + 0.45* S_{runtime}$ | $S_{quality}= 0.4* S_{pot}- S_{cst}$ |
|---|---|---|---|---|---|---|
| | $P_{filesize}$ | $P'_{filesize}$ | | | | |
| search_orig.java | 0.704KB | | 0 | 0 | 0 | 0 |
| search_ArraySplit.java | | 1.135 KB | 0.2 | 0.61 | 0.182 | 62.07 |
| search_Arrayfold.java | | 1.136 KB | 0 | 0.61 | 0.092 | 32.16 |
| search_Arrayflatten.java | | 1.223 KB | 0 | 0.75 | 0.113 | 21.49 |

Table 5. Obfuscation quality analysis

The quality of obfuscation is measured on a 100 point scale and the analysis Table 5 shows that the obfuscation quality of 'search_ArraySplit.java' is better compared to other codes.

**2 FUTURE WORKS**

The functionality for improving the strength of obfuscation of the code (say, in Figures 15, 17), by obscuring the index of setArray( ), getArray( ) method calls and hiding integers, is to be incorporated in the tool.